\documentclass[aps,prd,twocolumn,superscriptaddress,amsfont,amssymb,amsmath,nofootinbib,showpacs,balancelastpage]{revtex4-1}

\usepackage{graphicx,longtable,natbib,mathrsfs,xcolor,ulem,array,float}
\usepackage{txfonts,subfigure,aas_macros}

\newcommand{\bs}{\boldsymbol}

\newcommand{\Msun}{M_\odot}

\newcommand{\q}{\bs{q}}
\newcommand{\s}{\bs{s}}
\newcommand{\x}{\bs{x}}
\newcommand{\jo}{\bs{j}_0}
\newcommand{\dq}{\Delta{\bs{q}}}

\begin{document} 
\addtolength{\hoffset}{-0.525cm}
\addtolength{\textwidth}{1.05cm}
\title{Lagrangian space remapping and the angular momentum reconstruction from cosmic structures}

\author{Sijia Li}
\affiliation{Department of Astronomy, Xiamen University, Xiamen, Fujian 361005, China}
\author{Ming-Jie Sheng}
\affiliation{Department of Astronomy, Xiamen University, Xiamen, Fujian 361005, China}
\author{Haikun Li}
\affiliation{Department of Astronomy, Xiamen University, Xiamen, Fujian 361005, China}
\author{Hao-Ran Yu}\email{haoran@xmu.edu.cn}
\affiliation{Department of Astronomy, Xiamen University, Xiamen, Fujian 361005, China}

\date{\today}
%\date{Received \today; published -- 00, 0000}
 
\begin{abstract}
Large scale structures provide valuable information of the primordial perturbations that encode the secrets of the origin of the Universe. It is an essential step to map between observables and their {\it initial coordinates}, called {\it Lagrangian space}, from which primordial perturbations transfer their information to structures via linear theory. By using numerical simulations and state-of-the-art reconstruction techniques, we report the accuracy of estimating the Lagrangian coordinates of galaxies and galaxy clusters, represented by dark matter halos in various ranges of mass, and study the accuracy of this remapping on the angular momentum (spin) reconstruction. Our work shows that galaxy groups and clusters, represented by halos with mass $\gtrsim 10^{13}M_\odot$, can be accurately remapped to Lagrangian space, and their spin reconstruction errors are dominated by the reconstructed initial gravitational potential. For all mass ranges, the errors of Lagrangian remapping, as well as redshift space distortions, play subdominant roles in estimating their angular momenta. This study explains the low correlation level between observed galaxy spins and reconstructed cosmic initial conditions and illustrates the potential of using angular momenta of cosmic structures to improve the reconstruction of primordial perturbations.
\end{abstract}

\maketitle

\section{Introduction}\label{sec.intro}

The observed cosmic structures originate from the primordial perturbations.
The statistical properties of these perturbations enable us to examine cosmological models,
such as inflation, and constrain cosmological parameters therein.
The gravitational instability and structure formation link the primordial perturbations
and low-redshift observables together \citep{Peebles_1969,2005MNRAS.360L..82R,2021JCAP...06..024M}.
{In the early, linear period of structure formation, the matter overdensity $\delta(\bs x)$ in real space and $\delta(\bs k)$ in Fourier space grow proportional to the linear growth factor $D_+(a)$, where $a$ is the scale factor.} 
This ceases to be true when nonlinear structure formation takes place. On one hand, the transportation of matter changes the comoving coordinates of perturbations,
namely, from {Lagrangian space}\footnote{The Lagrangian space is defined as the ``{initial}, { comoving}" coordinates of mass elements in the picture of structure formation. In $N$-body simulations, $N$-body particles represent phase-space ``sheets'' 
{(\cite[]{2020moco.book.....D}, Chapter 12)}, and when the initial conditions of the simulation are set at sufficiently high redshift, their comoving coordinates represent Lagrangian space.} to Eulerian space, with a displacement field link between two spaces.
On the other hand, the overdensity no longer grows as $D_+(a)$.
Because of these two facts, the nice linear correlation fails on scales starting from
$k\simeq 0.1\,h\,{\rm Mpc}^{-1}$. Even by modern techniques \citep{2017PhRvD..95d3501Y,2017MNRAS.469.1968P}, the reconstructed
linear density fields can only correlate with the true initial perturbations on scales
$k\lesssim 1\,h\,{\rm Mpc}^{-1}$.
This is because we can only estimate the displacement field of the structure formation and 
undo this transportation effect on large scales.
The features of initial density perturbations are significantly lost upon shell crossing
and the virialization of dark matter halos and galaxies.

Meanwhile, the asymmetry of the Lagrangian clustering generates angular momenta, and
this vector mode \citep{1998_Turok_vector_mode} eventually encodes in the rotation of galaxies on scales of only 
tens of kiloparsec, possibly the smallest scale observable features that can correlate
with the cosmic primordial perturbations. 
This profound link comes from the tidal torque theory \citep{1970Afz.....6..581D,1984_white} description of the origin
of the protohalo angular momenta, linear and nonlinear structure formation, and the relations
between dark matter and baryons.
Since the initial tidal environment of the protohalos determines their initial angular momenta, 
and is finally correlated with the observed galaxy angular momenta (hereafter, we shorten the 
``angular momentum vectors'' to spins) \citep{2021_motloch_na}, these observables can also be used to 
reconstruct the distributions of the initial gravitational potential field \citep{2014ApJ...794...94W,2016ApJ...831..164W}.
Two essential steps are required to bridge the spin observable with the primordial perturbations.
First, how initial conditions around the protohalo evolve to the spin observable.
This is initially described by the tidal torque theory, the spin reconstruction method \citep{2020PhRvL.124j1302Y},
and confirmed in cosmological simulations \citep{2021PhRvD.103f3522W,2023ApJ...943..128S}, which will be briefly reviewed in Sec. \ref{sec.meth}.
Second, the transportation effects must be undone.
The galaxies and their spins are observed in real or redshift space, whereas their
protohalos are located in Lagrangian space, which is not observable. 
Remapping spin observables to Lagrangian coordinates is crucial for spin correlations and reconstructions.

In this work, we use modern reconstruction techniques to study the accuracy of the Lagrangian space remapping
and estimate their effects on Eulerian-Lagrangian spin correlation. We also extend the Lagrangian space spin 
reconstruction methods to Eulerian space and compare the halo spins with large scale structures. These enable 
us to have discussions regarding the understanding of spin correlations, initial condition reconstruction 
methods, and intrinsic alignments. 
The rest of the paper is organized as follows.
In Sec. \ref{sec.meth}, we review the analytical methods and describe the reconstructed simulations.
In Sec. \ref{sec.result}, we show the Lagrangian space remapping results and the
analysis of spin correlation. In Sec. \ref{sec.conclu}, we present {our} conclusion and discussions.

\section{Methods}\label{sec.meth}

\subsection{Spin evolution and reconstruction}

The production of halo spin is well described by the tidal torque theory.
For each dark matter halo, its protohalo occupies a certain region in Lagrangian space,
where each mass element of the protohalo exerts an initial acceleration and velocity
parallel to the gradient of the primordial gravitational potential $-\nabla\phi$.
Calculating the spin of the protohalo with respect to its {c.m.}
by expanding the velocity vector up to the first order leads to the tidal torque expression
$j_i^{({\rm T})}\propto\epsilon_{ijk}I_{jl}T_{lk}$. Here we have used the element expression of the
tensor equation; $j_i$ is the $i$th element of the spin vector $\bs j$, $I_{jl}$ is the moment of
inertia tensor of the protohalo, $T_{lk}\equiv\partial_l\partial_k \phi$ is the tidal tensor, and $\epsilon_{ijk}$ is the
Levi-Civita symbol. 

Numerical simulations show that (i) the true initial spins $j_i^{(q)}$
of protohalos [where the superscript {${(q)}$} denotes that the spin is in Lagrangian space 
$\bs q$] are accurately approximated by $j_i^{({\rm T})}$, and (ii) in Eulerian space, 
i.e., at low redshifts, the final halo spins $j_i$ are still highly correlated to the initial
directions $j_i^{(q)}$ and $j_i^{({\rm T})}$ \citep{2020PhRvL.124j1302Y,2021PhRvD.103f3522W}. The qualitative explanation of the fact that
nonlinear structure formation does not destroy the spin conservation that
(i) during linear epochs, the gravitational potential nearly remains constant,
so the tidal torque direction conserves; (ii) in nonlinear epochs, the tidal torque decays
due to the expansion of the Universe and the fact that halos shrink and become more round.

Hydrodynamic simulations further show that the baryonic matter in the halos closely traces
the dark matter in Lagrangian space in terms of locations, sizes, and shapes, resulting in
high spin correlations in both Lagrangian and Eulerian spaces between dark matter, gas, and stellar components \citep{2015ApJ...812...29T,2023ApJ...943..128S,2023arXiv231107969S}. Thus, the observable Galaxy spins are able to trace the properties
of initial perturbations in the protohalo/protogalaxy regions.
{Because of} the absence of shape (moment of inertia) information of protohalos and protogalaxies,
our previous works \cite{yu_2020prl,2021_motloch_na} presented the method of reconstructing spin fields by the interaction between tidal
fields on different scales, written as 
\begin{eqnarray}\label{eq.ttt}
  L^{\rm Rec}_i\propto\epsilon_{ijk}T_{jl}({\bs q},r_q)T_{lk}({\bs q},r_q{_{,+}}),
\end{eqnarray} 
where $T_{ij}({\bs q},r_q)$ denotes the tidal tensor field at ${\bs q}$, smoothed on scale
$r_q$, and $r_q{_{,+}}$ denotes a scale slightly larger than $r_q$.
The spin correlation and reconstruction are not affected by the assembly history of halos,
as long as the Lagrangian region of the total halo is considered. In another word,
the coordinates ${\bs q}$ correspond to the {c.m. of all protosubhalos} in the spin observable,
and the Lagrangian radius [Eq.(12.64) of {Ref.}\cite{2020moco.book.....D},  will also {be} introduced later] corresponds to the total halo mass. As
a result, the orbital angular momentum of a merging system will also be considered in
the tidal torque theory and the spin reconstruction.

\subsection{Density reconstruction}
In order to extract information of the primordial perturbations from spin observables,
we first need to maximize the correlation between spin reconstruction and observed galaxy spins.
Without using the spin information, a number of algorithms are developed to restore 
the primordial cosmic information from the {low-redshift} large scale structures,
including the direct Gaussianization transforms \citep{1992MNRAS.254..315W}, wavelet filters \cite{2011A&A...531A..75E}, {and} running reverse 
$N$-body simulations \citep{2018MNRAS.473.3351R}. Remarkably, with the development of modern computing power,
approximate primordial perturbations can be directly obtained after hundreds to thousands 
of iterations; examples include isobaric reconstruction \citep{2017ApJ...841L..29W}, solving {Monge-Amp\`ere}
equations \citep{2003A&A...406..393M}, and Hamiltonian {Markov-chain} methods \citep{2013ApJ...772...63W,2014ApJ...794...94W}. These methods are referred to as { ``density reconstruction''}, where, compared to the density fields at low redshifts,
the reconstructed initial density field is much better correlated to the true initial density field.
They significantly recover the Fisher information contained in the matter power spectrum \citep{2017MNRAS.469.1968P}, %[cite Pan qiaoyin]
and improve the cosmological {constraints}, such as the baryonic acoustic oscillations \citep{2017ApJ...841L..29W}. %[cite Wang Xin (BAO reconstruction)].
These reconstruction methods undo the large scale displacement of matter, such that 
on medium, {quasinonlinear} scales, the Fourier modes contain information about the primordial
perturbations.

\subsection{Reconstructed {simulations}}
We first run an ``original'' simulation, corresponding to the true
cosmic initial conditions, the underlying structure formation and the true properties of
dark matter halos. A flat $\Lambda$CDM cosmology is assumed, with 
cosmological parameters $\Omega_{\rm m}=0.258$, $\Omega_{\Lambda}=0.742$, 
$\sigma_8=0.807$, $h=0.72,$ and $n_s=1.0$ \citep{wmap5}. 
By using the cosmological $N$-body simulation code {\small{CUBE}} \citep{2018ApJS..237...24Y},
$N_p=400^3$ particles are initialized at redshift $z_{\rm init}=100$ using 
the Zel'dovich approximation \citep{1970A&A.....5...84Z}, 
in a cubic box $L=100\,{\rm Mpc}/h$ per side with periodic boundary conditions.
Then, the system is evolved to redshift $z=0$ with the particle-particle 
particle-mesh force calculation \citep{1981csup.book.....H}. 
At $z=0$, dark matter halos are identified using the {friends-of-friends} (FoF) method \citep{1985ApJ...292..371D}. 
The particle mass of the $N$-body particles is approximately $1.12\times10^9\Msun/h$,
%$1.3\times10^9\Msun$
and we consider dark matter halos with at least $100$ particles, or, equivalently halo mass
$\gtrsim 1.12\times10^{11}\Msun/h$.
The ``grid initial condition" is used where Lagrangian positions of particles
are placed at each center of the cell, in a $N_g=400^3$ mesh,
so it is straightforward to acquire their Lagrangian properties.

For this original simulation, based on its density fluctuations at redshift $z=0$,
we reconstruct an estimated initial condition, denoted as the ``reconstructed'' 
initial density fluctuations, by the {Hamiltonian Markov chain Monte carlo algorithm, with particle-mesh based forward simulation (the} ``HMC+PM'' method) \citep{2014ApJ...794...94W}, with resolution of $200$ grids per dimension. 
There are a number of other reconstruction methods {\citep{2023JCAP...06..062Q,2023JCAP...03..059M,2023MNRAS.520.6256S,2023MNRAS.523.6272C,2023arXiv230313056J}} that are
also able to obtain estimated initial conditions or displacement fields.
Specifically, the HMC+PM method was applied to the {Sloan Digital Sky Survey} 
galaxy surveys, and an estimation of the initial condition of a portion of the
real Universe is obtained, named ELUCID \citep{2016ApJ...831..164W}. By running cosmological simulations
using the ELUCID initial conditions, the simulated universe restores the observed
one on large scales. However, on small scales, most of the individual halos and 
galaxies in ELUCID {cannot} match the galaxies in the observed universe.
As a result, we {cannot} directly compare the halo/galaxy properties, such as
the angular momentum, object by object, between the simulation and observations.
By using the spin reconstruction method \citep{yu_2020prl,2021_motloch_na}, we estimated
the galaxy angular momenta directly from the initial condition field of ELUCID
and compared them with observations, obtaining a weak but significant correlation.

\begin{figure}[htbp]
  \centering
  \includegraphics[width=1\linewidth]{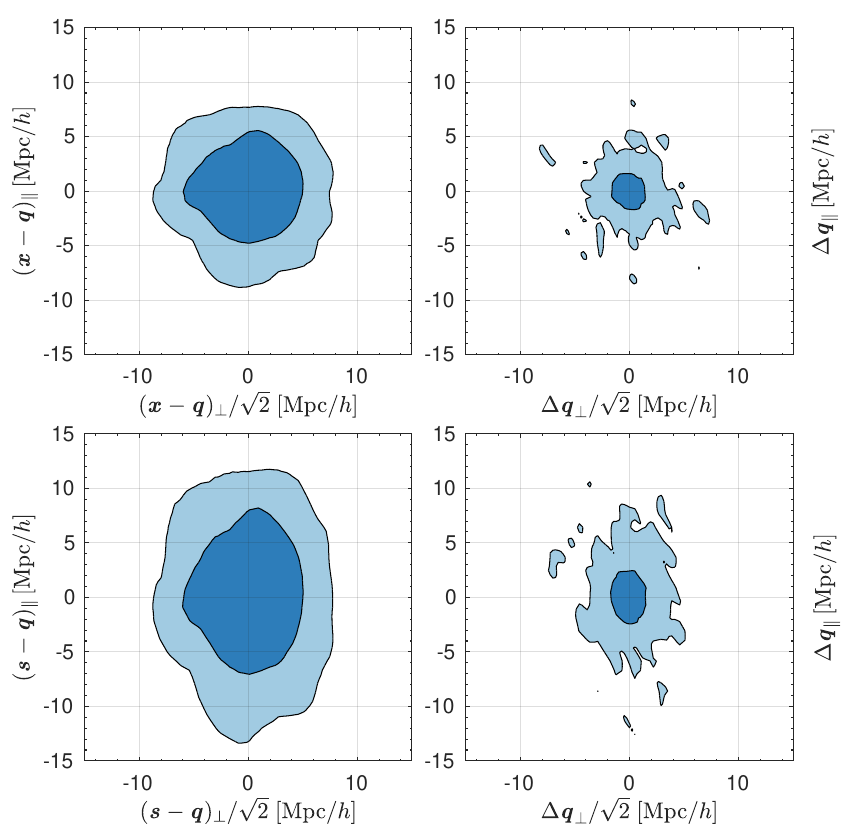}
   \caption{Lagrangian remapping accuracy of dark matter halos by the 
   reconstructed simulation. The horizontal and vertical axes represent
   the errors of the perpendicular and parallel components of the line of
   sight. The left/right {columns are} the errors before/after the
   reconstructed remapping, and the upper/lower {rows are} with/without
   redshift space distortions. The inner and outer contours contain {68\%} and 95\% of the halo population with mass $\geq 10^{12}\Msun/h $, respectively.} 
   \label{fig.1}
\end{figure}

With the reconstructed initial condition, we run a reconstructed simulation with 
$200^3$ particles in order to recover the Lagrangian properties of halos in the 
original simulation. We label the estimated properties or the values in the 
reconstructed simulation {with a} ``$\ \hat{}\ $''. The trajectories of the $N$-body 
particles in the reconstructed simulation provide a displacement field 
$\hat{\bs{\Psi}}(\hat{\bs{q}})\equiv \hat{\bs{x}}-\hat{\bs{q}}$, uniformly sampled 
at the Lagrangian positions $\hat{\bs{q}}$ of those particles, pointing to their 
final (Eulerian) positions $\hat{\bs{x}}$ at redshift $z=0$. Now we construct an {\it inverse} 
displacement field $\hat{\bs{\Phi}}(\hat{\bs{x}})\equiv\hat{\bs{q}}-\hat{\bs{x}}$ 
on a $200^3$ grid in order to remap halos back to their Lagrangian positions. However, 
the inverse displacement field is not uniformly sampled by $N$-body particles in 
Eulerian space; in the cosmic voids, there are many empty grids, and in the 
overdense regions, there are multiple particles falling in.
To avoid having this vector field multivalued on a certain grid, we calculate 
the inverse displacement vector by averaging over all particles that are eventually 
located in the grid. For empty grids, their values are undefined 
or assigned with zero vectors {or,} alternatively, assigned with the inverse 
displacement vector of the last leaving particle, scaled by the linear growth factor 
up to redshift $z=0$.
In practice, we find that how to deal with the empty grids does not affect the 
calculation of halo Lagrangian positions because all the halo positions in the 
original simulation $\bs{x}$ also correspond to positive overdensities in the 
reconstructed simulation. The inverse displacement field is then filtered by a 
Gaussian window function of scale $\sim1.5\,{\rm Mpc}/h$, which we find optimally reconstructs the Lagrangian properties in later sections. 
\begin{figure}
  \centering
    \includegraphics[width=0.95\linewidth]{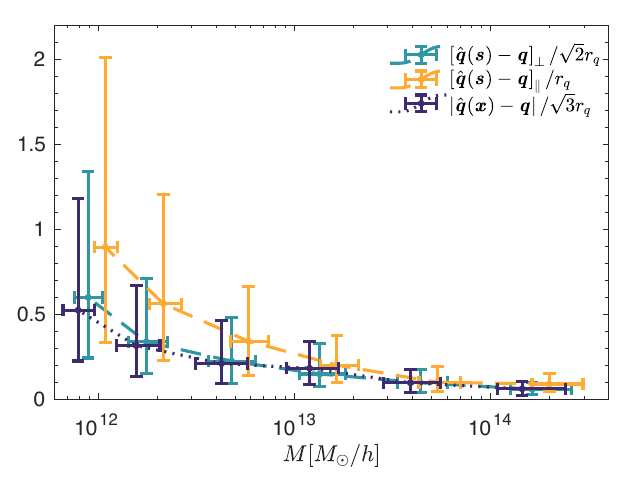}     
   \caption{Relative remapping errors $\Delta q / r_q$ as a function of halo mass, with and without RSD effect. When RSD is included, the error components are plotted parallel and perpendicular to {LOS}, respectively. The purple dotted curve shows $1/\sqrt{3}$ times of relative errors $\Delta q / r_q$, when remapped directly from real space (without RSD). The yellow and cyan dashed curves show parallel components ($\Delta q_{\parallel} / r_q$) and $1/ \sqrt{2}$ times the perpendicular components ($\Delta q_{\perp } / r_q$) of relative errors $\Delta q / r_q$, with RSD. All the points at the intersections show medians of data in each mass bin, and error bars are taken as $25\%$ and $75\%$ of the distribution. }
   \label{fig.2}
\end{figure}
For a given halo in Eulerian space $\bs{x}$, we interpolate its inverse displacement vector 
$\hat{\bs{\Phi}}(\bs{x})$ by the triangular shaped cloud method \citep{1981csup.book.....H} on the grid and 
find the estimated Lagrangian position $\hat{\bs{q}}(\bs{x})=\bs{x}+\hat{\bs{\Phi}}(\bs{x})$, with {redshift space distortions (RSDs)}.

More realistically, halos and galaxies are initially observed in redshift space. We also calculate the redshift space coordinates of halos $\bs{s}=\bs{x}+\bs{u}_\parallel/aH$ according to their {c.m.} peculiar velocities $\bs{u}$, where $\bs{u}_\parallel$ is the line of sight (LOS) component of $\bs{u}$, chosen here to be the second ($y$) axis of the simulation. Although methods like ELUCID can identify galaxy groups and correct the {RSDs} (both the Kaiser \citep{1987_kaiser} and finger of God effects {\citep{1972MNRAS.156P...1J,1978IAUS...79...31T}}), such that the halo positions are identified in real (Eulerian) space, here we also try to reconstruct the Lagrangian properties from the raw redshift space to study how RSD affects the spin reconstruction. In this case, the inverse displacement field is constructed on redshift space $\bs s$, i.e., $\hat{\bs{\Phi}}_s(\hat{\bs{s}})\equiv\hat{\bs{q}}-\hat{\bs{s}}$, and the Lagrangian position is reconstructed by $\hat{\bs{q}}(\bs{s})=\bs{s}+\hat{\bs{\Phi}}_s(\bs{s})$.

\begin{figure*}[htbp]
  \centering
   \includegraphics[width=0.95\linewidth]{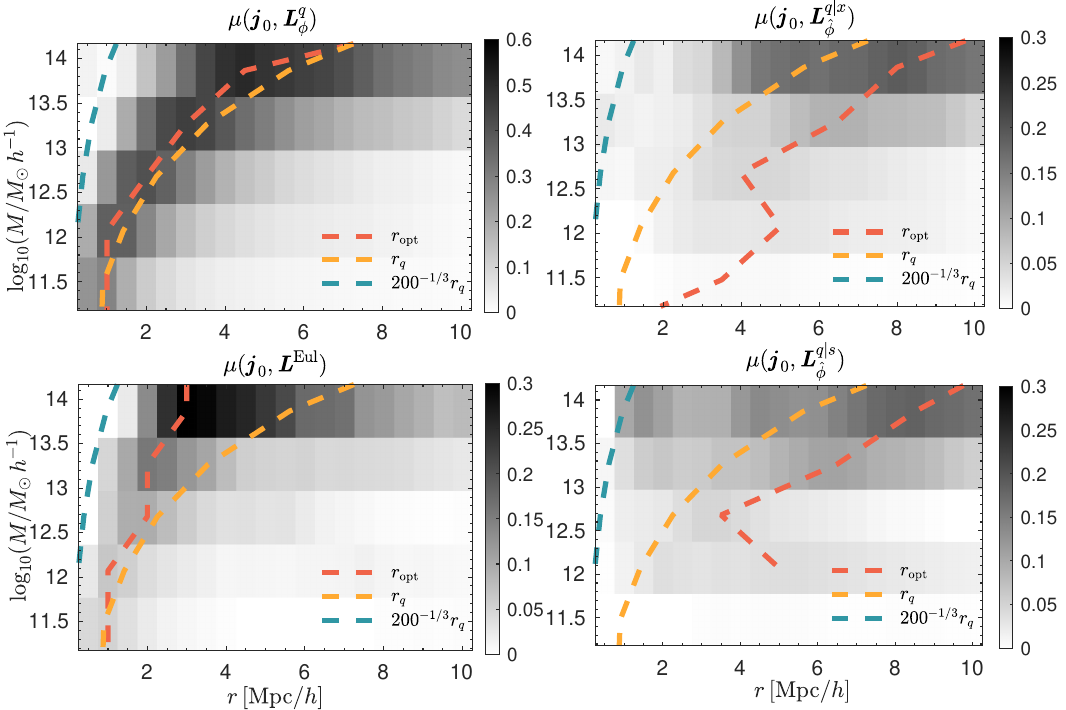}
   \caption{Optimal smoothing scale for spin reconstruction as a function of halo mass.
    The four panels show the cases of cross-correlating function 
    $\mu(\boldsymbol{j}_0,\boldsymbol{L}^{q}_{\phi})$ ({top left}), 
    $\mu(\boldsymbol{j}_0,\boldsymbol{L}^{\rm Eul})$ (bottom left), 
    $\mu(\boldsymbol{j}_0,\boldsymbol{L}_{\hat{\phi}}^{q|x})$ (top right) and 
    $\mu(\boldsymbol{j}_0,\boldsymbol{L}_{\hat{\phi}}^{q|s})$ (bottom right).
    The darker color of pixels means higher correlation and better reconstruction. 
    The red and yellow curves are optimal smoothing scale $r_{\rm opt}$ and 
    Lagrangian radius $r_q$.
    The cyan curve is $200^{-1/3}$ times the Lagrangian radius.
    }
   \label{fig.3}
\end{figure*}

\section{Results}\label{sec.result}
Here we quantify the results from the reconstructed simulation. 
First, we compare the remapped Lagrangian positions $\hat{\bs{q}}(\bs{x})$
and $\hat{\bs{q}}(\bs{s})$ with the true Lagrangian origins $\bs{q}$.
Here we select halos with mass greater than $10^{12}\Msun/h$.
In the left column of Fig. \ref{fig.1}, the contours show the errors 
parallel ($\parallel$, vertical axis) and perpendicular ($\perp$, horizontal)
to {LOS}, without the remapping process. The errors essentially come from the displacement of matter in the structure formation, and RSD brings additional errors in {the LOS} direction. By using the reconstructed simulation, the halo positions are remapped back to Lagrangian coordinates
with much better accuracy. The errors ($\Delta \bs{q}=\hat{\bs{q}}-\bs{q}$) are shown in the right column of Fig.\ref{fig.1}.
The $1 \sigma$ errors of the Lagrangian space remapping {are confirmed} to be $1.5 {\rm Mpc}/h$ in {the} perpendicular direction to {line of sight} and $2.5 {\rm Mpc}/h$ in line of sight, which is much lower compared {to} errors without Lagrangian remapping, where the errors are $3.2 {\rm Mpc}/h$ and $4.8 {\rm Mpc}/h$.
When the Lagrangian positions are remapped directly from the redshift space,
the errors in {the LOS} direction are slightly larger.
The errors after remapping are mainly caused by the loss of small scale information in reconstructing the initial conditions.

The {low-redshift} halos and galaxies can reflect the properties of their Lagrangian
regions, which occupy a certain comoving volume. The errors of halo Lagrangian 
positions should be compared with the scale of those Lagrangian regions.
Dark matter halos are the largest virialized objects in the {Universe}, and
their Lagrangian counterparts, the protohalos in Lagrangian space, are
typically compact and simply connected. This is confirmed by our
previous work \citep{2023ApJ...943..128S}. It is natural to use the Lagrangian equivalent radius
$r_q\equiv(2M G/\Omega_{\rm m}H_0^2)^{-1/3}$ to characterize this scale,
where $M$, $G$, $\Omega_{\rm m}$, {and} $H_0$ are the halo mass, Newton's constant, 
current matter density parameter, and the Hubble constant, respectively.
Figure. \ref{fig.2} shows the errors of Lagrangian remapping relative to each halo's
equivalent Lagrangian radius $r_q$. For the RSD contaminated case, these relative
error vectors are plotted by LOS and transverse (perpendicular to LOS) components,
and we divide all halos into six mass bins to show the mass dependence of the relative error.
The centers of the error bars show the median of the distribution in each mass bin, 
while the left/lower and right/upper error bar boundaries show the $25$ and $75$
percentiles of the distribution.
The general trend is that the relative remapping error $\Delta q/r_q$
is decreasing with the increase of halo mass. 
In particular, for halo masses greater than $10^{13}\Msun/h$, the remapped
Lagrangian positions are close to true values, with errors much smaller 
than the halo Lagrangian radii $r_q$.
When the halo masses approach $10^{12}\Msun/h$ or smaller, the errors are comparable
to $r_q$. For the RSD contaminated case, we also plot the LOS and transverse
error components separately. Note that, if the error is isotropic, statistically
$\Delta q_\parallel=\Delta q_\perp/\sqrt{2}$.
The results show that halos with mass $\gtrsim 10^{13}\Msun/h$ are uncontaminated by RSD,
but for less massive halos, RSD brings additional uncertainties on $q_\parallel$
by about a factor of {2}.

\begin{figure}[htbp]
  \centering 
   \includegraphics[width=1\linewidth]{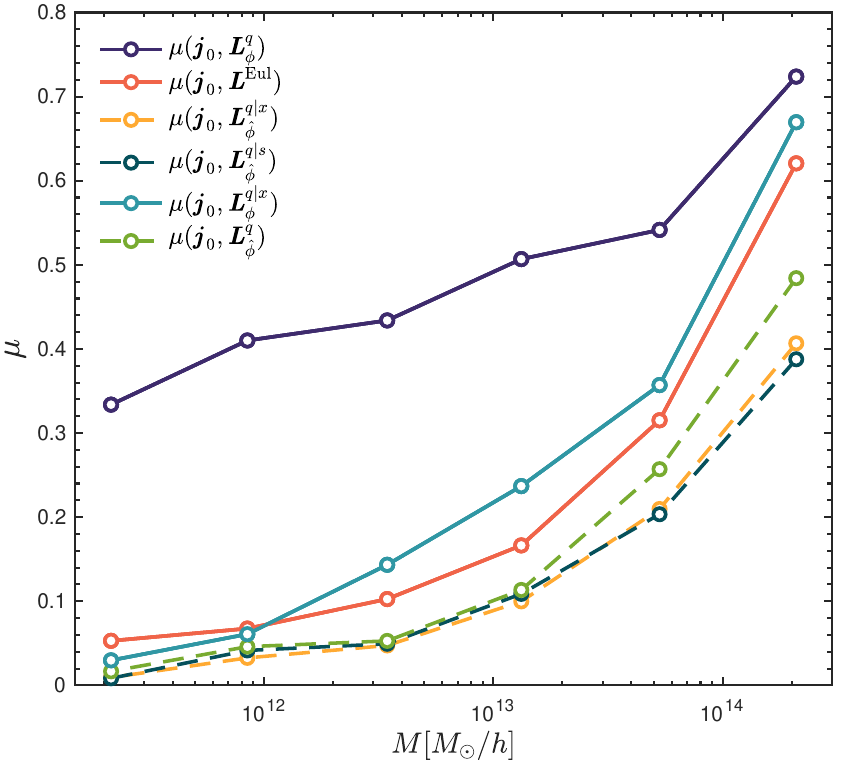}
   \caption{Maximally achievable cross-correlation of spin reconstruction of halos in different mass ranges.
    The spin reconstructed from $\boldsymbol{q}$ space ($\boldsymbol{L}^{q}_{\phi}$) and $\boldsymbol{x}$ space ($\boldsymbol{L}^{\rm Eul}$) in the original simulation is cross-correlated with $\boldsymbol{j}_0$, and the results are shown by the purple and red solid curves, respectively. 
    The cross-correlations between the spin reconstructed from remapped coordinates $\boldsymbol{L}_{\hat{\phi}}^{q|x}$ and $\boldsymbol{L}_{\hat{\phi}}^{q|s}$ in the reconstructed simulation and $\boldsymbol{j}_0$ are illustrated
    by the yellow and dark cyan dashed curves. 
    With or without the remapping process or the reconstructed initial condition are also plotted for comparison by cyan and green curves.
    Circles at the curves mean the maximum reachable cross-correlation at each halo bin.
    }
   \label{fig.crosscorrelation}
\end{figure}

Next, we use the remapped Lagrangian positions and reconstructed primordial fluctuations 
to estimate the halo spin directions. First, 
in the top left panel of Fig. \ref{fig.3}, we directly use Eq.(\ref{eq.ttt}), with the true Lagrangian positions
and true tidal fields $T_{ij}\equiv\partial_i\partial_j\phi$ derived from the
original initial potential field $\phi$. The {cross-correlations} 
$\mu(\boldsymbol{j}_0,\boldsymbol{L}^{q}_{\phi})$ are quantified by the cosine of the angle
between the two angular momentum vectors. The gray scale plot illustrates $\mu$
as a function of the smoothing scale $r$ and the halo mass $M$.
The red dashed curve shows the optimal smoothing scales $r_{\rm opt}$ which maximize
$\mu$ in the given halo mass bins, while in this panel $r_{\rm opt}$ is compared with 
the Lagrangian equivalent radii $r_q$ (yellow dashed curve) and approximate halo scales 
$200^{-1/3}r_q$, which is the radius of a sphere whose mean density is 200 times the critical density of the Universe (cyan dashed curve). 
It is clear and expected that these two scales are similar
for all halo mass bins.\footnote{
  The trace difference in contrast to \cite{yu_2020prl} is due to our usage of
  FoF halos instead of spherical overdensity (SO) halos.
} The correlation $\mu$ is high for all halo masses, while
showing the slightly increasing function of $M$.

Next, we study the spin correlation using the reconstructed initial 
conditions (reconstructed potential $\hat{\phi}$) and the remapped Lagrangian coordinates.
On the superscript, we use the symbol ``$q|x$'' to denote that the Lagrangian coordinate is estimated given the halo position $\boldsymbol{x}$, i.e., $\hat{\bs{q}}(\bs{x})$. Similarly, the superscript ``$q|s$'' stands for the estimated Lagrangian coordinate $\hat{\bs{q}}(\bs{s})$ given redshift position $\boldsymbol{s}$. On the subscript, we denote which potential field ($\phi$, $\hat\phi$, or $\phi_0$) is used for the spin reconstruction.
For example, $\bs{L}_{\hat{\phi}}^{q|x}$ is the spin field reconstructed by $\hat{\phi}$ and $\hat{\bs{q}}(\bs{x})$,
$\bs{L}_{{\phi}}^{q}$ means that the spin field is reconstructed by the real initial condition, and the halo spin is correlated with its real Lagrangian position in the field. $\bs{L}^{\rm Eul}$ is the spin reconstructed by $\phi_0$ and $\bs{x}$, i.e., $\bs{L}_{{\phi_0}}^{x}$.

Using remapped Lagrangian positions [$\hat{\bs{q}}(\bs{x})$ or $\hat{\bs{q}}(\bs{s})$] and the reconstructed initial 
conditions ($\hat{\phi}$), the spin field is correlated with the low redshift halo spin vectors, yielding the mean cross correlation 
$\mu(\boldsymbol{j}_0,\boldsymbol{L}_{\hat{\phi}}^{q|x})$ and $\mu(\boldsymbol{j}_0,\boldsymbol{L}_{\hat{\phi}}^{q|s})$ in each halo mass bin.
They are plotted in the right two panels of Fig. \ref{fig.3}.
Compared to the idealized linear spin reconstruction $\bs{L}_{{\phi}}^{q}$ (top left panel of Fig.\ref{fig.3}, where real $\phi$ as well as real $\bs q$ is used), the correlations decrease significantly across all mass ranges, by $30\% \sim 40\%$.
For halos with mass $\gtrsim 10^{13}\Msun/h$, corresponding to galaxy clusters,
the correlation is still greater than $20\%$, while for less massive halos
(about $10^{12}\Msun/h$), the correlation drops to about $5\%$.
Another feature is that the optimal smoothing scale $r_{\rm opt}$ is shifted to 
larger scales. 
This result directly explains the low correlation between observed galaxy spins and reconstructed cosmic initial conditions (ELUCID) in \cite{2021_motloch_na} and the relatively large smoothing scale used there.
Physically, the reconstructed initial condition loses correlation with the true initial condition on small scales,
and thus the cosmic information is lost in this regime. {In addition}, the uncertainties
on $\bs q$ also favor a larger $r_{\rm opt}$, such that the larger scale
still-correlated tidal interactions can be included in the spin estimation.
Comparing the two right panels of Fig.\ref{fig.3}, we find that RSD has
an insignificant influence on the spin estimation.

It is instructive to investigate whether the tidal-torque-based spin estimator
also works in Eulerian space. Replacing the primordial tidal tensors $T_{ij}(\bs q)$ 
with Eulerian values $T^{\rm Eul}_{ij}(\bs x)\equiv\partial_i\partial_j\phi_0$, 
where $\phi_0$ is the Eulerian gravitational potential at $z=0$,
and using the Eulerian positions $\bs x$, the Eulerian-space-based spin estimator
is written as
\begin{eqnarray}\label{eq.ttt_x}
  L^{\rm Eul}_i\propto\epsilon_{ijk}T^{\rm Eul}_{jl}({\bs x},r)T^{\rm Eul}_{lk}({\bs x},r{_{,+}}),
\end{eqnarray} 
where $r$ is again the arbitrary smoothing scale.
The result in the bottom left panel of Fig.\ref{fig.3} shows that there are also correlation signals
from the Eulerian-space-based spin estimator. 
The correlations are found in all mass ranges but are significantly lower than in the Lagrangian space (top left panel). The other feature is that the optimal
smoothing scale $r_{\rm opt}$ is somewhat located between $r_q$ and $200^{-1/3}r_q$.
This may lead to the fact that the low-redshift cosmic structures, especially on scales
slightly larger than the halo scale, do have an influence on the halo angular momentum 
distribution and galaxy intrinsic alignments.

Finally, we summarize the maximally achievable spin correlation $\mu_{\rm max}$ as a function of halo mass $M$.
 We calculate the cross-correlations between the $\bs{j}_{0}$ and the $\bs{L}$ reconstructed with several pairs of initial conditions and coordinates; we refer the reader to the Appendix~\ref{appendix} for specific definitions. In Fig.\ref{fig.crosscorrelation}, the top curve is obtained by the true initial condition $\phi$ and the true Lagrangian position $\q$, which is the idealized scenario given our spin reconstruction method. Its deviation from perfect correlation ($\mu=1$) comes from the approximations used in the tidal torque theory, spin reconstruction, and the nonlinear evolution. In practice, what we can use in the real space is the estimated $\hat{\phi}$ and remapped $\hat{\bs{q}}(\bs{s})$. Even $\hat{\bs{q}}(\bs{x})$ can only be derived with the permission of redshift distortion correction from $\s$. 
Comparing the two curves $\mu({\bs j}_0,{\bs L}_{\hat\phi}^{q|x})$ and $\mu({\bs j}_0,{\bs L}_{\hat\phi}^{q|s})$, we find that whether Lagrangian positions are remapped from real or redshift space has little effect on the upper limit of the cross-correlation--the RSD effect is well corrected in the remapping process.

The decreasing of the cross-correlation comes from the combination of two effects--the error of remapping and the error of initial gravitational potential. In order to see which one is the dominant effect on the decorrelation, we plot the two curves $\mu({\bs j}_0,{\bs L}_{\phi}^{q|x})$ and $\mu({\bs j}_0,{\bs L}_{\hat\phi}^q)$, i.e., using the true $\phi$ but estimated $\bs q$, and the true $\bs q$ but estimated $\phi$. It appears that the error on the initial potential $\phi$ has more severe effects on spin reconstruction--the accuracy of the initial condition is more important than the accuracy of the remapping process in spin reconstruction. This shows the potential of better reconstruction of $\hat\phi$, given fixed $\hat{\bs{q}}(\bs{x})$, if additional spin information of galaxies/halos are used. 

An additional curve $\mu({\bs j}_0,{\bs L}^{\rm Eul})$ is also plotted in Fig.\ref{fig.crosscorrelation} to show the maximally achievable correlation between low-redshift halo spins and low-redshift large scale structure. Although the Eulerian spin reconstruction is comparable to Lagrangian ones (with estimated quantities), it cannot be straightforwardly used to reconstruct the initial perturbations. Nevertheless, it is closely related to galaxy/halo intrinsic alignments, and is additionally parity odd. We leave this connection and physical explanation to future works.

\section{conclusion and discussions}\label{sec.conclu}
 
  In this paper, by using the reconstructed simulations, we study the accuracy of Lagrangian space remapping--the initial comoving coordinates of virialized halo-galaxy systems are obtained. Inspired by the tidal torque theory, the halo angular momentum vector (spin) is reconstructed in Lagrangian space (spin reconstruction), and is used to assess the remapping quality. The main results are summarized as follows.
  
\begin{itemize}
    \item The remapping process can remarkably reduce the errors induced by the displacement of matter in the structure formation, which is a crucial step in spin reconstruction. Statistically, the remapping errors are small compared to the Lagrangian radii of the protohalos, especially for massive ones.
    \item The correlation between halo spin and the reconstructed spin obtained from remapped coordinates and reconstructed initial gravitational potential drops compared to the idealized situation. The latter uses the true Lagrangian coordinates. The deviation significantly increases in low mass ranges, which explains the low correlation between observed galaxy spins and the spin reconstruction (and the optimal smoothing scale) in the real Universe. RSD effect increases the remapping error parallel to the line of sight, but induces minor effects in the final spin reconstruction.
    \item The accuracy of the reconstructed initial gravitational potential is the main factor affecting the performance of the spin reconstruction, especially on small scales, while the remapping is relatively 
    accurate.\vspace{3mm}
    \item {Our results indicate that more massive halos exhibit better remapping precision and stronger spin correlations.
    This is because larger mass ranges correspond to larger, more linear scales in Lagrangian space, and the linear behavior of the cosmic displacement field. Compared to the work \citep{2021_motloch_na} using spiral galaxies tracing less massive halos, more massive galaxy groups/clusters, and even filaments \citep{2022PhRvD.105f3540S}, show the potential in improving the mean spin correlation {\it per object}, and explore the clustering on linear scales. Although limited by the abundance of these more massive objects, any cosmological constraints should, in principle, be optimized by utilizing all available cosmological information.}
\end{itemize}

Remapping the low-redshift observed objects back to Lagrangian space has profound importance in cosmology. Using the spin evolution of halo-galaxy systems as an example, the low-redshift spin vectors are reliable tracers of tidal environment of the protohalos. Observing the former infers the latter and can further help with obtaining the cosmic initial conditions. The displacement field relates these quantities at different coordinates--Eulerian (real or redshift space) and Lagrangian, and in this work we have shown that Eulerian space can be reliably remapped to Lagrangian space, can whose errors do not erase the spin correlation. The lowering of spin correlation is mainly due to the poor reconstruction of initial gravitational potential on small scales. Nevertheless, the initial gravitational potential is obtained by density reconstruction, which used only galaxy positions. By additionally using galaxy spins (especially the directions), the initial gravitational potential can be better constrained, and along with it, the displacement field, and equivalently the Lagrangian remapping can be further improved. Iterating these steps can help obtain a more accurate initial state of our Universe, and we leave it for our subsequent works.

\section*{ACKNOWLEDGMENTS}
{We thank the anonymous referee for valuable suggestions.}
This work is supported by the National Science Foundation of China Grant No. 12173030.
We thank ELUCID Collaboration for providing reconstructed simulations. 
The calculations of this work were performed on the workstation of cosmological sciences, Department of Astronomy, Xiamen University.

\bibliographystyle{h-physrev3}
\bibliography{sijia_ref}

\begin{thebibliography}{10}

\bibitem{Peebles_1969}
P.~J.~E. {Peebles},
\newblock \apj {\bf 155}, 393 (1969).

\bibitem{2005MNRAS.360L..82R}
C.~D. {Rimes} and A.~J.~S. {Hamilton},
\newblock \mnras {\bf 360}, L82 (2005), astro-ph/0502081.

\bibitem{2021JCAP...06..024M}
M.~{McQuinn},
\newblock \jcap {\bf 2021}, 024 (2021), 2008.12312.

\bibitem{2020moco.book.....D}
S.~{Dodelson} and F.~{Schmidt},
\newblock {\em {Modern Cosmology}} (, 2020).

\bibitem{2017PhRvD..95d3501Y}
H.-R. {Yu}, U.-L. {Pen}, and H.-M. {Zhu},
\newblock \prd {\bf 95}, 043501 (2017), 1610.07112.

\bibitem{2017MNRAS.469.1968P}
Q.~{Pan}, U.-L. {Pen}, D.~{Inman}, and H.-R. {Yu},
\newblock \mnras {\bf 469}, 1968 (2017), 1611.10013.

\bibitem{1998_Turok_vector_mode}
N.~{Turok}, U.-L. {Pen}, and U.~{Seljak},
\newblock \prd {\bf 58}, 023506 (1998), astro-ph/9706250.

\bibitem{1970Afz.....6..581D}
A.~G. {Doroshkevich},
\newblock Astrofizika {\bf 6}, 581 (1970).

\bibitem{1984_white}
S.~D.~M. {White},
\newblock \apj {\bf 286}, 38 (1984).

\bibitem{2021_motloch_na}
P.~{Motloch}, H.-R. {Yu}, U.-L. {Pen}, and Y.~{Xie},
\newblock Nature Astronomy {\bf 5}, 283 (2021), 2003.04800.

\bibitem{2014ApJ...794...94W}
H.~{Wang}, H.~J. {Mo}, X.~{Yang}, Y.~P. {Jing}, and W.~P. {Lin},
\newblock \apj {\bf 794}, 94 (2014), 1407.3451.

\bibitem{2016ApJ...831..164W}
H.~{Wang} {\em et~al.},
\newblock \apj {\bf 831}, 164 (2016), 1608.01763.

\bibitem{2020PhRvL.124j1302Y}
H.-R. {Yu} {\em et~al.},
\newblock \prl {\bf 124}, 101302 (2020), 1904.01029.

\bibitem{2021PhRvD.103f3522W}
Q.~{Wu}, H.-R. {Yu}, S.~{Liao}, and M.~{Du},
\newblock \prd {\bf 103}, 063522 (2021), 2011.03893.

\bibitem{2023ApJ...943..128S}
M.-J. {Sheng} {\em et~al.},
\newblock \apj {\bf 943}, 128 (2023), 2210.04203.

\bibitem{2015ApJ...812...29T}
A.~F. {Teklu} {\em et~al.},
\newblock \apj {\bf 812}, 29 (2015), 1503.03501.

\bibitem{2023arXiv231107969S}
M.-J. {Sheng} {\em et~al.},
\newblock arXiv e-prints , arXiv:2311.07969 (2023), 2311.07969.

\bibitem{yu_2020prl}
H.-R. {Yu} {\em et~al.},
\newblock \prl {\bf 124}, 101302 (2020), 1904.01029.

\bibitem{1992MNRAS.254..315W}
D.~H. {Weinberg},
\newblock \mnras {\bf 254}, 315 (1992).

\bibitem{2011A&A...531A..75E}
J.~{Einasto} {\em et~al.},
\newblock \aap {\bf 531}, A75 (2011), 1012.3550.

\bibitem{2018MNRAS.473.3351R}
H.~{Rein} and D.~{Tamayo},
\newblock \mnras {\bf 473}, 3351 (2018), 1704.07715.

\bibitem{2017ApJ...841L..29W}
X.~{Wang} {\em et~al.},
\newblock \apjl {\bf 841}, L29 (2017), 1703.09742.

\bibitem{2003A&A...406..393M}
R.~{Mohayaee}, U.~{Frisch}, S.~{Matarrese}, and A.~{Sobolevskii},
\newblock \aap {\bf 406}, 393 (2003), astro-ph/0301641.

\bibitem{2013ApJ...772...63W}
H.~{Wang}, H.~J. {Mo}, X.~{Yang}, and F.~C. {van den Bosch},
\newblock \apj {\bf 772}, 63 (2013), 1301.1348.

\bibitem{wmap5}
J.~{Dunkley} {\em et~al.},
\newblock \apjs {\bf 180}, 306 (2009), 0803.0586.

\bibitem{2018ApJS..237...24Y}
H.-R. {Yu}, U.-L. {Pen}, and X.~{Wang},
\newblock \apjs {\bf 237}, 24 (2018), 1712.06121.

\bibitem{1970A&A.....5...84Z}
Y.~B. {Zel'dovich},
\newblock \aap {\bf 5}, 84 (1970).

\bibitem{1981csup.book.....H}
R.~W. {Hockney} and J.~W. {Eastwood},
\newblock {\em {Computer Simulation Using Particles}} (, 1981).

\bibitem{1985ApJ...292..371D}
M.~{Davis}, G.~{Efstathiou}, C.~S. {Frenk}, and S.~D.~M. {White},
\newblock \apj {\bf 292}, 371 (1985).

\bibitem{2023JCAP...06..062Q}
F.~{Qin}, D.~{Parkinson}, S.~E. {Hong}, and C.~G. {Sabiu},
\newblock \jcap {\bf 2023}, 062 (2023), 2302.02087.

\bibitem{2023JCAP...03..059M}
C.~{Modi}, Y.~{Li}, and D.~{Blei},
\newblock \jcap {\bf 2023}, 059 (2023), 2206.15433.

\bibitem{2023MNRAS.520.6256S}
C.~J. {Shallue} and D.~J. {Eisenstein},
\newblock \mnras {\bf 520}, 6256 (2023), 2207.12511.

\bibitem{2023MNRAS.523.6272C}
X.~{Chen}, F.~{Zhu}, S.~{Gaines}, and N.~{Padmanabhan},
\newblock \mnras {\bf 523}, 6272 (2023), 2306.10538.

\bibitem{2023arXiv230313056J}
V.~{Jindal}, A.~{Liang}, A.~{Singh}, S.~{Ho}, and D.~{Jamieson},
\newblock arXiv e-prints , arXiv:2303.13056 (2023), 2303.13056.

\bibitem{1987_kaiser}
N.~{Kaiser},
\newblock \mnras {\bf 227}, 1 (1987).

\bibitem{1972MNRAS.156P...1J}
J.~C. {Jackson},
\newblock \mnras {\bf 156}, 1P (1972), 0810.3908.

\bibitem{1978IAUS...79...31T}
R.~B. {Tully} and J.~R. {Fisher},
\newblock {Nearby Small Groups of Galaxies},
\newblock in {\em Large Scale Structures in the Universe}, edited by M.~S.
  {Longair} and J.~{Einasto}, volume~79, p.~31, 1978.

\bibitem{2022PhRvD.105f3540S}
M.-J. {Sheng} {\em et~al.},
\newblock \prd {\bf 105}, 063540 (2022), 2110.15512.

\end{thebibliography}

\newpage
\appendix
\section{Variable description}\label{appendix}
We present descriptions of the physical variables used in this paper in Table \ref{Tab1}.

\begin{table}[b]
\caption{Variables used in this paper.}\label{Tab1}
\centering
 \begin{tabular}{ll} 
 \hline
Variables & Description \\
\hline
$\q$ & True Lagrangian space coordinates \\
$\x$ & Eulerian space  coordinates   \\
$\s$ & Redshift space  coordinates   \\
$\hat{\q}$ &  $\q$ in reconstructed simulations \\
$\hat{\bs\Psi}(\q)$ & Displacement field    \\
$\hat{\bs\Phi}(\x)$, $\hat{\bs\Phi}(\s)$ & Inverse displacement field  \\
$\jo$ & Eulerian spin vector \\
$\dq$ &  $\q-\hat{\q}$\\
$r_q$ &  Lagrangian radius\\
$r_{\rm opt}$ & Optimal smoothing scale\\
$\phi$ & Gravitation field in Lagrangian space\\
$\phi_0$ & Gravitation field in Eulerian space\\
$\hat{\phi}$ & Reconstructed initial gravitation field\\
$\epsilon_{ijk}$ & Levi-Civita symbol\\
$I_{ij}$ & Momentum of inertia tensor of the protohalo\\
$T_{ij}$ & Tidal tensor field\\ 
$\bs{L}^{\rm Eul}$ &  Spin reconstructed by $\phi_0$ and $\x$\\
$\bs{L}^{q}_{\phi}$ & Spin reconstructed by the $\phi$ and $\q$\\
$\bs{L}^{q}_{\hat{\phi}}$ & Spin reconstructed by the $\hat{\phi}$ and $\q$\\ 
%\hline
$\bs{L}^{q|x}_{\hat{\phi}}$ & Spin reconstructed by the $\hat{\phi}$ and $\hat{\bs{q}}(\bs{x})$\\
%\hline
$\bs{L}^{q|s}_{\hat{\phi}}$ & Spin reconstructed by the $\hat{\phi}$ and $\hat{\bs{q}}(\bs{s})$\\
%\hline
$\bs{L}^{q|x}_{\phi}$ & Spin reconstructed by the $\phi$ and $\hat{\bs{q}}(\bs{x})$\\ 
  \hline
 \end{tabular}
\end{table}

\end{document}